\documentclass[eqsecnum,twocolumn,showpacs,preprintnumbers,amssymb,pra]{revtex4}
\usepackage{graphicx}
\usepackage{dcolumn}
\usepackage{amsthm,amsmath}

\begin{document}

\preprint{\today}

\title{Revisiting Nuclear Quadrupole Moments in $^{39-41}$K Isotopes}

\author{Yashpal Singh \footnote{Email: yashpal@prl.res.in}, D. K. Nandy and B. K. Sahoo \footnote{Email: bijaya@prl.res.in}}

\affiliation{Theoretical Physics Division, Physical Research Laboratory,
Navrangpura, Ahmedabad - 380009, India}

\begin{abstract}
Nuclear quadrupole moments ($Q$s) in three isotopes of potassium (K) with atomic 
mass numbers 39, 40 and 41 are evaluated more precisely in this work. The $Q$
value of $^{39}$K is determined to be 0.0614(6) $b$ by combining the available experimental
result of the electric quadrupole hyperfine structure constant ($B$) with 
our calculated $B/Q$ result of its $4P_{3/2}$ state. Furthermore combining
this $Q$ value with the measured ratios $Q$($ ^{40}$K)$/Q$($ ^{39}$K) and 
$Q$($ ^{41}$K)$/Q$($ ^{39}$K), we obtain $Q$($ ^{40}$K)$=-0.0764(10) \ b$ and
$Q$($ ^{41}$K)$=0.0747(10) \ b$, respectively. These results disagree with the 
recently quoted standard values in the nuclear data table within the given 
uncertainties. The calculations are carried out by employing the relativistic
coupled-cluster theory at the singles, doubles and involving important valence
triples approximation. The accuracies of the calculated $B/Q$ results can be
viewed on the basis of comparison between our calculated magnetic dipole hyperfine
structure constants ($A$s) with their corresponding measurements for many low-lying
states. Both $A$ and $B$ results in few more excited states are presented for the 
first time.
\end{abstract}


\maketitle

\section{Introduction}
Potassium (K) atom has three naturally abundant isotopes with atomic mass numbers 39, 40 and 41.
Using the modern femtosecond laser frequency combs and polarization quantum-beat techniques,
high precision measurements of hyperfine structure constants in $4P$ and $3D$ states are carried
out \cite{Yei,flake}. Also, a number of measurements of these quantities were carried out
in the ground and other states long ago using the atomic beam magnetic resonance and level crossing techniques (e.g. see
review article by Arimondo {\it et al.} \cite{Arimondo}). Theoretical studies of these quantities
are of immense interest to atomic physicists to test the accuracies of the wave functions in
the nuclear region \cite{sahoo1, sahoo2, nataraj}. However, theoretical evaluation of these
quantities require atomic calculations and nuclear moments \cite{sahoo1, sahoo2, sahoo, bsahoo}.
Nuclear magnetic moments ($\mu$s) of the above K isotopes are known very precisely and the reported 
results from various studies matches reasonably well with each other \cite{stone}. On the other hand,
the reported nuclear quadrupole moments ($Q$s) from various works on these isotopes differ significantly.
 For example, $Q$ value of $^{39}$K is reported as 0.07(2) $b$ \cite{series}, 0.049(4) $b$ \cite{Arimondo},
0.0601(15) $b$ \cite{olsen} and 0.0585 $b$ within one percent error \cite{kello}. The latest result,
0.0585(6) $b$, is now considered as the standard $Q$ value for $^{39}$K \cite{stone,webelement}.
Accurate knowledge of $Q$ values of these isotopes are useful in many applications. These information
are interesting in order to test the potential of nuclear models \cite{baum, neyens}, acquiring 
information about local symmetry \cite{feld}, to find out asymmetry parameters in nuclei 
\cite{errico,asahi}, for studying the Mossbauer spectroscopy for the structural 
determination of the element containing solid state compounds \cite{martinez} etc.

In this paper, we analyze the electric quadrupole hyperfine structure constants for many states in K 
and report precise $Q$ values of its above mentioned isotopes. As discussed later, we find the new 
values to be larger than the considered standard values in the literature. Atomic wave functions are
calculated using relativistic coupled-cluster (RCC) method in the Fock-space representation and matrix 
elements of the hyperfine interaction Hamiltonians are estimated using these wave functions in the 
considered atom. 

The rest of the paper is organized as follows: In the next section, we present briefly the theory
of hyperfine structure in an atomic system and the single particle matrix elements of the interaction
Hamiltonians which are used to evaluate the hyperfine structure constants. In Sec III, we explain the 
RCC method little elaborately for the calculation of atomic wave functions. Then, we present the results 
and their discussions before summarizing the work. Unless stated otherwise, we use atomic unit (au) 
throughout this paper. 

\section{Theory of Hyperfine Structure}
The hyperfine structures of energy levels in an atom arise due to the interaction between electron angular
momenta with the nuclear spin. Details of this theory is given by C. Schwartz in a classic paper 
\cite{schwartz}. Mathematically, the hyperfine interaction Hamiltonian is given in a general form as 
non-central interaction between electrons and the nucleus in terms of tensor operators as
\begin{equation}
 H_{hfs}=\displaystyle \sum _k T_e^{(k)} \cdotp T_n^{(k)},
\end{equation}
where $T_e^{(k)}$ and $T_n^{(k)} $ are the spherical tensor operators of rank $k$ in the space of 
electronic and nuclear coordinates, respectively. In the first order perturbation theory, the 
hyperfine interaction energy $W_F$ of hyperfine state $|F;IJ\rangle$ with total angular momentum $F=I+J$ 
for $I$ and $J$ being the nuclear spin and electronic angular momentum of the associated fine structure 
state $|J,M_J\rangle$, respectively, taking up to $k=2$ is given by 
\begin{equation}
 W_F=\frac{1}{2}A R +B \frac{\frac{3}{2}R(R+1)-2I(I+1)J(J+1)}{2I(2I-1)2J(2J-1)},
\end{equation}
with $R=F(F+1)-I(I+1)-J(J+1)$, and $A$ and $B$ are known as the magnetic dipole and electric quadrupole hyperfine 
structure constant for $k=1$ and $k=2$, respectively. The advantage of expressing the change in energy in this 
form is it separates out the electronic and nuclear factors for which the calculations can be carried out in a 
simple approach.
Here $A$ and $B$ are given by \cite{schwartz, cheng}
\begin{equation}
  A=\mu_N g_I \frac{\langle J|| T_e^{(1)}||J\rangle}{\sqrt{J(J+1)(2J+1)}},
\end{equation}
and
\begin{equation}
 B=Q \left \{ \frac{8J(2J-1)}{(2J+1)(2J+2)(2J+3)} \right\} \langle J|| T_e^{(2)}||J\rangle.
\end{equation}
In the above expressions, $\mu_N$ and $g_I=\mu/I$ are the nuclear magneton and gyromagnetic ratio, respectively. 
Since our intention is to verify accuracies of $Q$ values, we estimate $B/Q$ results in this work.

The reduced matrix elements of the electronic spherical tensor operators, $T_e^{(k)}=\sum t_e^{(k)}$, in terms of
single orbitals are given by \cite{schwartz, cheng}
\begin{eqnarray}
 \nonumber \langle \kappa_f||t_e^{(1)}|\kappa_i \rangle= -(\kappa_f + \kappa_i) \langle -\kappa_f||C^{(1)} || \kappa_i \rangle \\ \int _0 ^{\infty}
dr \frac{P_f Q_i + Q_f P_i}{r^2}
\label{heqn1}
\end{eqnarray}
and
\begin{equation}
  \langle \kappa_f||t_e^{(2)}||\kappa_i \rangle= -\langle \kappa_f||C^{(2)} || \kappa_i \rangle\int _0 ^{\infty}
\label{heqn2}
dr \frac{P_f Q_i + Q_f P_i}{r^3}
\end{equation}
where $\kappa_i$ and $P_i$ ($Q_i$) are the relativistic angular momentum quantum number and large (small)
component of Dirac spinor for the corresponding orbital $i$, respectively. The reduced matrix elements of
Racah tensors ($C^{(k)}$) are given by \cite{edmond}
\begin{eqnarray}
\nonumber  \langle \kappa_f||C^{(k)}||\kappa_i \rangle= (-1)^{j_f+1/2} \sqrt{(2j_f+1)(2j_i+1)}  \\
\times
\left \{ \begin{matrix}
 j_f & k & j_i\\ 
 1/2& 0 & 1/2 \\
\end{matrix} \right \}
\pi(\ell_f,k,\ell_i)
\end{eqnarray}
with the angular momentum selection rule $\pi(\ell_f,k,\ell_i)=1$ when $\ell_f+k+\ell_i= \text{even}$ for the orbital
angular momentum $\ell_f$ and $\ell_i$; otherwise it is zero.

\section{Methods for Calculations}

\subsection{Single particle orbital generation}
Accurate generation of atomic orbitals in the nuclear region is very important for the present study. We
consider here Gaussian type of orbitals (GTOs) which provide natural description of relativistic wave-functions
within nucleus \cite{aerts,visser,mc} as basis to construct the mean-field orbitals in the Dirac(Hartree)-Fock (DF)
approach. Kinetic balanced condition between the large and small components of Dirac spinor are imposed to ensure
correct non-relativistic behavior of the orbitals \cite{mc, stanton}. GTOs to construct an orbital at a particular
location $r_i$ are defined as 
\begin{equation}
 F^{L(S)}(r_i)= \sum_k \mathcal{N}_k^{L(S)} r_i^{l+1} e^{-\eta_k r_i^2}
\end{equation}
where $L(S)$ represents for large (small) component, $k$ denotes number of
GTOs, $\mathcal{N}$ correspond to normalization factor for each GTO and $\eta_k$ is an arbitrary parameter which
has to be chosen suitably for orbitals from different $\ell$ symmetries. To get more flexibility in optimization of our basis sets, we use 
the even tempering condition by defining two more parameters $\zeta$ and $\nu$ as
\begin{equation}
 \eta_k=\zeta \nu^{k-1}.
\end{equation}
The radial grid points $r_i$ are defined as
\begin{eqnarray}
 r_i = r_0 [e^{h(i-1)} - 1],
\end{eqnarray}
with $r_0$ is the starting radial function taken inside the nucleus to be $2 \times 10^{-6}$ at which the wave functions become
finite and $h$ is a step size which is defined by taking maximum radial function $r_{max}$ as $150.0$ au
and total grid points $1000$.

We have considered 40 GTOs for each $l$ symmetry orbitals and the considered $\zeta$ and $\nu$ are
given in Table \ref{tab1} for different $l$ values. Due to limitation over computational resources and
negligible contributions from the high lying virtual orbitals, we have taken up to 24 orbitals from 
$s$, $p$, $d$ symmetries and 17 orbitals from $f$, $g$ symmetries to construct active space for RCC calculations.
\begin{table}
\caption{\label{tab1}
Used $\zeta$ and $\nu$ parameters for different '$\ell$' symmetries to construct GTOs.
}
\begin{ruledtabular}
\begin{tabular}{cccccc}
 & \textrm{s}&\textrm{p}&\textrm{d}&\textrm{f}&\textrm{g}\\
\colrule
$\zeta$    & 0.0002    & 0.0004   &   0.0003 & 0.0005   & 0.0004  \\
$\nu$       & 1.917     & 1.79     &   1.77   & 1.76     & 1.75 \\
\end{tabular}
\end{ruledtabular}
\end{table}

Also, the orbitals are generated by accounting the finite size of the nucleus assuming a two-parameter Fermi-nuclear-charge
distribution given by
\begin{equation}
\rho(r_i) = \frac {\rho_0} {1 + e^{(r_i-c)/a}},
\end{equation}
where $\rho_0$ is the density for the point nuclei, $c$ and $a$ are the half-charge radius and skin thickness of the
nucleus. These parameters are chosen as
\begin{equation}
a = 2.3/4(ln3)
\end{equation}
and
\begin{equation}
c = \sqrt{ \frac{5}{3} r_{rms}^2 - \frac{7}{3} a^2 \pi^2},
\label{eqn31}
\end{equation}
where $r_{rms}$ is the root mean square radius of the corresponding nuclei which is taken as $3.61$ fm \cite{ragavan}.

\subsection{Calculation of atomic wave functions}
\begin{figure}
 \includegraphics[width=7cm]{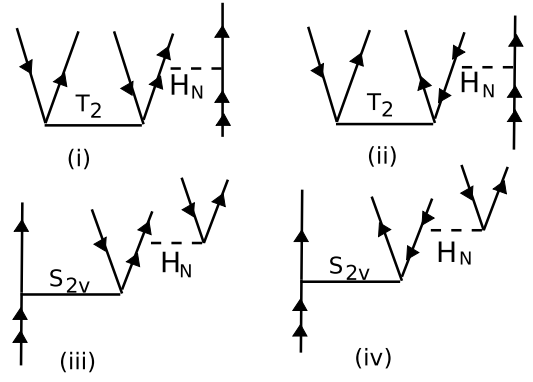}
\caption{\label{fig1} Typical Goldstone diagrams representing leading-order triple excitations over the
CCSD method. Double arrows in the diagrams represents valence electron $v$, and the lines with upward (downward)
arrows represents particle (hole) orbitals.}
\end{figure}

To calculate matrix elements of the hyperfine interaction Hamiltonian, we use the RCC method where we define atomic wave functions for the
considered states with valence orbital denoted by $v$ as \cite{lindgren, mukherjee}
\begin{equation}
 |\Psi_v \rangle=e^T\{1+S_v\}|\Phi_v\rangle,
 \label{cceqn}
\end{equation}
where the DF wave function $|\Phi_v\rangle$ is constructed as $|\Phi_v\rangle=a_v^{\dagger}
|\Phi_0 \rangle$ with $|\Phi_0 \rangle$ is the DF wave function for the closed-shell
configuration $[3p^6]$ in the considered K atom. In the above expression, $T$ and $S_v$ are the
excitation operators that accounts for core and core-valence correlations to all orders, respectively.
Since K is a small size atom, hence correlation effects among electrons are expected to be less. Therefore
role of the higher order configurations in determining atomic wave-function could be negligible. On the 
other-hand consideration of these configurations are computationally very expensive. Owing to this fact 
we account only all possible single and double configuration excitations to all orders (known as CCSD 
method) by expressing the above operators in the Fock space representation as
\begin{eqnarray}
\nonumber  T=T_1+T_2 &=& \displaystyle \sum _{a,p} a_p^{\dagger} a_a t_a^p + \frac{1}{4} 
\displaystyle \sum_{ab,pq}  a_p^{\dagger} a_q^{\dagger} a_b a_a t_{ab}^{pq} \\
\\
\nonumber S_v=S_{1v}+S_{2v}&=& \displaystyle \sum _{a,p} a_p^{\dagger} a_v s_v^p + \frac{1}{2} 
\displaystyle \sum_{ab,pq}  a_p^{\dagger} a_q^{\dagger} a_b a_v s_{vb}^{pq}\\
\end{eqnarray}
where the ($a,b,c \cdots$), ($p,q,r \cdots$) and ($v$) subscripts of the second quantized operators 
represents core (hole), particle (virtual) and valance orbitals, respectively. However expanding 
Eq. (\ref{cceqn}) using these CCSD operators to all non-linear terms give rises contributions from 
higher excitations. We determine the above $t$ and $s_v$ coefficients which correspond to the excitation 
amplitudes using the following equations
\begin{equation}
\langle \Phi ^L|\{\widehat{H_N e^T}\}|\Phi_0 \rangle=0
\end{equation}
and 
\begin{eqnarray}
 \nonumber \langle \Phi_v ^L|\{\widehat{H_N e^T}\} S_v|\Phi_v \rangle=-
\langle \Phi_v ^L|\{\widehat{H_N e^T}\}|\Phi_v \rangle \\
+\langle \Phi_v^L|S_v|`\Phi_v \rangle \Delta E_v,
\label{opcc}
\end{eqnarray}
with the superscript L(=1,2) representing the single and double excited configurations from the corresponding
DF states, the wide-hat symbol denotes the linked terms, $\Delta E_v$ is the attachment energy of the valence 
electron $v$ and $H_N$ denotes the normal ordering atomic Dirac-Coulomb Hamiltonian $H$ which is taken as
\begin{eqnarray}
 H = \sum_i [c \mbox{\boldmath$\alpha$}_i \cdotp \mbox{\boldmath$p$}_i + (\beta_i -1)c^2] + \sum_{i \ge j } \frac{1}{r_{ij}}, 
\end{eqnarray}
where $\mbox{\boldmath$\alpha$}$ and $\beta$ are usual Dirac matrices, $c$ is the velocity of light. $\Delta E_v$ is evaluated by
\begin{equation}
 \Delta E_v=\langle \Phi_v|\{\widehat{H_N e^T}\}\{1+ S_v\}|\Phi_v \rangle.
 \label{trip}
\end{equation}
To improve the quality of energy and calculation of wave functions due to the dominant triple
excitations containing the valence orbital, we define a perturbation operator $S_{3v}$ by 
contracting $H_N$ with $T_2$ and $S_{2v}$ operators as
\begin{equation}
 S_{3v}(s_{vbc}^{pqr})=\frac{\widehat{H_N T_2} + \widehat{H_N S_{2v}}}{\epsilon_p+\epsilon_q+\epsilon_r
-\epsilon_b - \epsilon_c - \epsilon_v},
\end{equation}
with $s_{vbc}^{pqr}$ correspond to excitation amplitudes and $\epsilon_i$ is the DF energy of the 
electron in the $i^{th}$ orbital. This operator is considered as a 
part of $S_v$ operator in Eq. (\ref{trip}) to get additional contribution to $\Delta E_v$. Since
$\Delta E_v$ is involved in Eq. (\ref{opcc}), we solve both the equations simultaneously in an
iterative procedure. This approach is generally referred as CCSD(T) method \cite{kaldor}. The
diagrammatic representation of these excitations are shown in Fig. \ref{fig1}.  

The expectation values due to the hyperfine interaction operators have been evaluated using our RCC method by
\begin{eqnarray}
 \nonumber \langle T_e^{(k)} \rangle &=& \frac{\langle \Psi_v |T_e^{(k)}|\Psi_v \rangle}{\langle \Psi_v |\Psi_v \rangle} \\
\nonumber &=&\frac{\langle \Phi_v|\{1+S_v^{\dagger}\} \overline T_e^{(k)} \{1+S_v\}|\Phi_v\rangle}
{\{1+S_v^{\dagger}\} \overline N_0 \{1+S_v\}}\\ \nonumber
&=&\frac{\langle \Phi_v|\{1+S_{1v}^{\dagger}+S_{2v}^{\dagger}\} \overline T_e^{(k)} \{1+S_{1v}+S_{2v}\}|\Phi_v\rangle}
{\{1+S_{1v}^{\dagger}+S_{2v}^{\dagger}\} \overline N_0 \{1+S_{1v}+S_{2v}\}} \\
\label{prpeqn}
\end{eqnarray}
where $\overline T_e^{(k)}=(e^{T^{\dagger}} T_e^{(k)} e^T)$ and $\overline N_0=e^{T{^\dagger}}e^T$.
Generally, both $\overline T_e^{(k)}$ and $\overline N_0$ in our RCC approach are non-terminating series. These terms are
terminated keeping terms minimum up to fourth order in perturbation. Description of this procedure
has been given in the previous works \cite{sahoo-sur,sahoo-gopa,sahoo-das}. Contributions from normalization 
of the wave functions ($Norm$) are estimated explicitly in the following way
\begin{equation}
 Norm=\langle \Psi_v|T_e^{(k)}|\Psi_v \rangle \left \{ \frac{1}{1+N_v} -1\right\}
\end{equation}
where $N_v= \{1+S_{1v}^{\dagger}+S_{2v}^{\dagger}\} \overline N_0 \{1+S_{1v}+S_{2v}\}$.

\begin{table*}                                                         
\caption{\label{tab2}
Comparison of calculated and available experimental $A$ results in $^{39-41}K$ (in MHz). Theoretical $A$s in different isotopes 
are evaluated using the calculated CCSD(T) results of $A/g_I$ in $^{39}$K and their respective experimental $g_I$ values.
Uncertainties estimated from the calculations are given in parentheses of our results.}
\begin{ruledtabular}
\begin{tabular}{ccccccc}
\textrm{State} &  \multicolumn{3}{c}{\textrm{This work}} & \multicolumn{3}{c}{\textrm{Experiment}} \\
  \cline{2-4} \cline{5-7} \\   &  $^{39}$K  & $^{40}$K & $^{41}$K  &  $^{39}$K  & $^{40}$K & $^{41}$K \\
\colrule \\
$4S_{1/2}$    &     229.6(2.0)& $ -$285.5(2.5) &     126.0(1.1)   &  230.8598601(3)\cite{Arimondo}&    $-$285.7308(24)\cite{Arimondo}& 127.0069352(6) \cite{Arimondo}\\
$4P_{1/2}$    &      27.4(5)  & $ -$34.04(62)  &     15.02(27)    &  27.775(42)\cite{flake}       &   $-$34.523(25)\cite{flake}    & 15.245(42) \cite{flake}        \\
              &               &                &                  &  28.85(30) \cite{Arimondo}    &   $-$34.49(11)\cite{bendali}   & 15.19(21) \cite{bendali}\\
              &               &                &                  &  27.80(15) \cite{bendali}     &                              & 15.1(8) \cite{touchard}\\
              &               &                &                  &  27.5(4) \cite{touchard}      &                              &\\
              &               &                &                  &  28.859(15)\cite{banerjee}    &                              &\\
$4P_{3/2}$    &      5.9(2)   & $ -$7.35(37)   &     3.25(16)     &  6.093(25)\cite{flake}        &   $-$7.585(10)\cite{flake}     & 3.363(25)\cite{flake}  \\
              &               &                &                  &  6.06(8) \cite{Arimondo}      &   $-$7.48(6) \cite{bendali}    & 3.40(8) \cite{ney}\\
              &               &                &                  &  6.00(10) \cite{schmieder}    &   $-$7.59(6) \cite{ney2}       & 3.325(15) \cite{sieradzan}\\
              &               &                &                  &  6.13(5) \cite{ney}           &                  &\\
$3D_{3/2}$    &      1.0(2)   & $ -$1.25(25)   &    0.55(11)      &  0.96(4)\cite{Yei}            &   1.07(2)\cite{Yei}& 0.55(3) \cite{Yei}\\
$3D_{5/2}$    &     $-$0.57(5)& $  $0.711(62)  &   $-$0.314(27)   &   0.62(4)\cite{Yei}           &   0.71(4)\cite{Yei}& 0.40(2) \cite{Yei}\\
$4D_{3/2}$    &      0.686(4) & $ -$0.853(5)   &    0.377(2)      &                               &                  &         \\
$4D_{5/2}$    &   $-$0.332(2) & $  $0.413(2)   &   $-$0.182(1)    &                               &                  &          \\
$5S_{1/2}$    &     55.0(1.0) & $ -$68.4(1.2)  &     30.18(55)    &  55.50(60)\cite{Arimondo}     &                  &         \\
$5P_{1/2}$    &      8.8(5)   & $ -$10.98(62)  &     4.84(27)     &  9.02(17)\cite{Arimondo}      &                  &         \\
$5P_{3/2}$    &      1.9(2)   & $ -$2.37(25)   &     1.04(11)     &  1.969(13) \cite{Arimondo}    &  $-$2.45(2) \cite{Arimondo}& 1.08(2) \cite{Arimondo}\\
              &               &                &                  &   1.95(5) \cite{schmieder}    &                  &\\
$5D_{3/2}$    &      0.39(5)  & $ -$0.489(62)  &    0.22(27)      &  0.44(10)\cite{Arimondo}      &                  &         \\
$5D_{5/2}$    &    $-$0.171(7)& $  $0.213(9)   &   $-$0.094(4)    &  $\pm$0.24(7)\cite{Arimondo}  &                  &         \\
$6S_{1/2}$    &      21.1(6)  & $ -$26.24(75)  &     11.58(33)    &  21.81(18)\cite{Arimondo}     &                  &  12.03(40) \cite{Arimondo}\\
$6P_{1/2}$    &      3.9(4)   & $ -$4.87(50)   &     2.15(22)     &  4.05(7)\cite{Arimondo}       &                  &          \\
$6P_{3/2}$    &      0.9(2)   & $ -$1.17(25)   &    0.51(11)      &  0.886(8)\cite{Arimondo}      &                  &         \\
$6D_{3/2}$    &      0.24(5)  & $ -$0.297(62)  &    0.131(27)     &  0.25(3)\cite{glodz}          &                  &         \\
              &               &                &                  & $\pm$ 0.2(2) \cite{Arimondo}  &                  &          \\
$6D_{5/2}$    &  $ -$0.112(5) & $  $0.139(6)   &   $-$0.061(3)    & $-$0.12(4)\cite{glodz}        &                  &         \\
              &               &                &                  & $\pm$ 0.10(10) \cite{Arimondo}&                  &          \\
$7S_{1/2}$    &   10.3(5)     & $ -$12.83(62)  &     5.66(27)     & 10.79(5)\cite{Arimondo}       &                  &         \\
$7P_{1/2}$    &   2.1(3)      & $ -$2.61(37)   &     1.15(16)     & $\pm$ 2.18(5)\cite{Arimondo}  &                  &         \\
$7P_{3/2}$    &   0.50(3)     & $ -$0.62(37)   &    0.28(16)      &                               &                  &         \\
$7D_{3/2}$    &   0.15(1)     & $ -$0.19(12)   &    0.083(5)      &                               &                  &          \\
$7D_{5/2}$    &  $-$0.068(5)  & $  $0.085(6)   &   $-$0.037(3)    &                               &                  &         \\
$8S_{1/2}$    &   5.8(2)      & $ -$7.27(25)   &     3.21(11)     & 5.99(8)\cite{Arimondo}        &                  &         \\
$8P_{1/2}$    &     1.3(1)    & $ -$1.56(12)   &    0.69(55)      &                               &                  &         \\
$8P_{3/2}$    &     0.301(5)  & $ -$0.374(6)   &    0.165(3)      &                               &                  &         \\
$8D_{3/2}$    &     0.101(2)  & $ -$0.126(2)   &    0.055(11)     &                               &                  &         \\
$8D_{5/2}$    &   $-$0.040(1) & $  $0.050(1)   &   $-$0.022(1)    &                               &                  &         \\
$9S_{1/2}$    &     3.4(3)    & $ -$4.25(37)   &     1.88(16)     &                               &                  &          \\
$9P_{1/2}$    &     1.0(1)    & $ -$1.19(12)   &    0.53(55)      &                               &                  &         \\
$9P_{3/2}$    &      0.230(4) & $ -$0.286(5)   &    0.126(2)      &                               &                  &         \\
$9D_{3/2}$    &     0.334(4)  & $ -$0.415(5)   &    0.183(2)      &                               &                  &         \\
$9D_{5/2}$    &    $-$0.148(2)& $  $0.184(2)   &   $-$0.081(1)    &                               &                  &         \\
$10S_{1/2}$   &   2.2(3)      & $ -$2.68(37)   &     1.18(16)     &   2.41(5)\cite{Arimondo}      &                  &         \\
\end{tabular}                                                                                                              
\end{ruledtabular}                                                                                                         
\end{table*}

\begin{table*}[t]
\caption{\label{tab3}
Comparison between different theoretical results of $A$ in $^{39}$K (in MHz). }
\begin{ruledtabular}
\begin{tabular}{ccccccc}

\textrm{State}&  \multicolumn{3}{c}{Others \cite{safronova}} & \multicolumn{3}{c}{This Work} \\
   \cline{2-4} \cline{5-7}   &    DF   &  SD & SDpT &  DF & CCSD  &  CCSD(T) \\
\colrule
$4S_{1/2}$      & 146.91 & 237.40   & 228.57  & 146.794 & 229.573      & 229.556  \\
$4P_{1/2}$      & 16.616 & 28.689   & 27.662  & 16.616  & 27.247       &  27.371  \\
$4P_{3/2}$      & 3.233  & 6.213    & 5.989   & 3.234   & 5.886        &   5.913      \\
$3D_{3/2}$      & 0.447  & 0.983    & 1.111   & 0.447   & 1.006        &   1.003       \\
$3D_{5/2}$      & 0.192  &$-$0.535  &$-$0.639 & 0.192   & $-$0.574     &  $-$0.572     \\
$4D_{3/2}$      & 0.281  & 0.678    &         & 0.281   & 0.690        &   0.686       \\
$4D_{5/2}$      & 0.120  &$-$0.307  &         & 0.120   & $-$0.334     &   $-$0.332    \\
$5S_{1/2}$      & 38.877 & 56.102   & 54.817  & 38.847  & 55.070       &  54.981        \\  
$5P_{1/2}$      & 5.735  & 9.202    & 8.949   & 5.735   & 8.755        &   8.827        \\
$5P_{3/2}$      & 1.117  & 1.988    & 1.932   & 1.117   & 1.887        &   1.903        \\
$5D_{3/2}$      & 0.168  & 0.409    &         & 0.168   & 0.396        &   0.393        \\
$5D_{5/2}$      & 0.072  &$ -$0.167 &         & 0.072   & $-$0.173     & $-$0.171       \\
$6S_{1/2}$      & 15.759 & 22.025   & 21.609  & 15.105  & 21.167       &   21.10         \\  
$6P_{1/2}$      & 2.629  & 4.066    & 4.014   & 2.629   & 3.874        &   3.918         \\
$6P_{3/2}$      & 0.512  & 0.874    & 0.866   & 0.512   & 0.928        &   0.937         \\
$6D_{3/2}$      & 0.105  & 0.253    &         & 0.104   & 0.241        &   0.239         \\
$6D_{5/2}$      & 0.0448 &$-$0.0975 &       & 0.045   & $-$0.113     &$ -$0.112        \\
$7S_{1/2}$      & 7.900  & 10.876   & 10.690  & 7.894   & 10.363       & 10.317           \\
$7P_{1/2}$      & 1.417  & 2.191    & 2.140   & 1.417   & 2.066        & 2.095            \\
$7P_{3/2}$      & 0.276  & 0.473    & 0.462   & 0.276   & 0.495        & 0.502         \\
$7D_{3/2}$      & 0.0685 & 0.1644   &         & 0.067   & 0.154        & 0.152        \\
$7D_{5/2}$      & 0.0293 &$-$0.0611 &         & 0.028   & $-$0.069     &$-$0.068      \\
$8S_{1/2}$      & 4.511  & 6.156    & 6.057   & 4.536   & 5.880        & 5.847            \\  
$8P_{1/2}$      &        &          &         & 0.855   & 1.236        &   1.257          \\
$8P_{3/2}$      &        &          &         & 0.166   & 0.296        &   0.301          \\
$8D_{3/2}$      &        &          &         & 0.048   & 0.102        &   0.101          \\
$8D_{5/2}$      &        &          &         & 0.019   & $-$0.040     & $-$0.040        \\
$9S_{1/2}$      & 2.814  & 3.818    & 3.759   & 2.685   & 3.444        &   3.420         \\
$9P_{1/2}$      &        &          &         & 0.695   & 0.945        &   0.958          \\  
$9P_{3/2}$      &        &          &         & 0.136   & 0.227        &    0.230         \\
$9D_{3/2}$      &        &          &         & 0.189   & 0.338        &   0.334          \\
$9D_{5/2}$      &        &          &         & 0.083   & $-$0.151     &  $-$0.148        \\
$10S_{1/2}$     & 1.871  & 2.529    & 2.491   & 1.878   & 2.171        & 2.154            \\
\end{tabular}                                                                                                          
\end{ruledtabular}                                                                                                     
\end{table*}

\section{Results and Discussions}
Our aim is to obtain $B/Q$ values more accurately in different states of K atom so that they can be combined
with the available precise experimental results for $B$ to estimate $Q$. In order to verify the accuracies of 
$B/Q$ results from our calculations, it would be felicitous to test the accuracies of the wave functions
in the nuclear region. Owing to the fact that our calculation procedure deals with many numerical 
computations at different stages, along with that it correlates with higher excitations configurations 
indirectly, hence it would be very difficult to estimate uncertainties from the used numerical methods 
and approximations taken at the level of excitations. With the intention of verifying accuracies of the 
wave functions in the nuclear region, we have calculated $A$ for many states in K. Assuming that the 
anomalous effects due to different nuclear sizes in all the considered isotopes are very small, we 
evaluate $A/g_I$ in $^{39}$K and determine $A$ values for the corresponding isotopes using their 
respective $g_I$ values. We have used experimental values of $g_I (^{39}K)=0.2609772$, 
$g_I (^{40}K)=-0.324525$ and $g_I (^{41}K)=0.1432467$ \cite{stone} to estimate these quantities ignoring 
their uncertainties as they will not meddle the results within the reported uncertainties. 
Both the calculated and experimental results are compared in Table \ref{tab2}. We estimate 
uncertainties in our calculations by considering incompleteness of basis functions, contributions 
from the inactive orbitals in the RCC method, higher order excitation levels and from the neglected 
terms in the non-truncative series of Eq. (\ref{prpeqn}). The upper limit to these uncertainties are 
given in the parentheses of the above table. Clearly, our estimated uncertainties are fair enough to 
compare with the available experimental results and our assumption for neglecting the anomalous effects 
for different isotopes seem to be reasonable.

\begin{table*}
\caption{\label{tab4}
Calculated $B/Q$ values (in MHz/$b$) from our DF, CCSD and CCSD(T) methods and accurately known experimental $B$ results (in MHz) in $^{39}$K and $^{41}$K. 
The most precise $B$ results are given in bold fonts and $Q$ values are estimated by combining these results with the calculated $B/Q$ values 
in both the isotopes.}
\begin{ruledtabular}
\begin{tabular}{cccccccc}
\textrm{State}& &\multicolumn{2}{c}{Theory}& \textrm{Experiment} & \textrm{$Q(^{39}K)$}& \textrm{Experiment}& \textrm{$Q(^{41}K)$} \\
          &   $DF$    &   $CCSD$&   $CCSD(T)$ & $B (^{39}K)$ & & $B (^{41}K)$&\\
\colrule
$4P_{3/2}$& 23.000& 44.392 &  44.6(5)&  {\bf 2.786(71)} \cite{flake}    & \underline{0.0625(17)} & 3.351(71) \cite{flake}      &      \\
          &       &        &         &  2.72(12)  \cite{banerjee}       &                        & 3.34(24)  \cite{ney}        &     \\
          &       &        &         &  2.83(13)  \cite{risberg}        &                        & {\bf 3.320(23)} \cite{sieradzan}  &  \underline{0.0744(10)}       \\        
          &       &        &         &  2.9(2)  \cite{schmieder}        &                        &                             &       \\
$5P_{3/2}$& 7.866 & 13.833 &  13.9(4)&  {\bf 0.870(22)} \cite{Arimondo} & 0.0626(22)             & {\bf1.06(4)}   \cite{ney}        &  0.0763(36)   \\
          &       &        &         &  0.92(10) \cite{schmieder}       &                        &                             &  \\
$6P_{3/2}$& 3.340 & 6.285  &  6.4(5) &  {\bf 0.370(15)} \cite{Arimondo} & 0.0578(51)             &                             &      \\
\end{tabular}
\end{ruledtabular}
\end{table*}

\begin{table*}
\caption{\label{tab5}
Contributions from different RCC terms to $A$ (in MHz) and $B/Q$ (in MHz/$b$), whereas c.c. stands for complex conjugation.
}
\begin{ruledtabular}
\begin{tabular}{c cc cc cc cc cc}
\textrm{State}& \multicolumn{2}{c}{$\overline{O}$} & \multicolumn{2}{c}{$\overline{O} S_{1v}~ + ~c.c.$} &
 \multicolumn{2}{c}{$\overline{O} S_{2v}~ + ~c.c.$} &\multicolumn{2}{c}{\textrm{Others}} & \multicolumn{2}{c}{\textrm{$Norm$}} \\
\cline{2-3} \cline{4-5} \cline{6-7}  \cline{8-9} \cline{10-11}\\ & $A$ & $B/Q$ & $A$ & $B/Q$ & $A$ & $B/Q$ & $A$ & $B/Q$ & $A$ & $B/Q$ \\
\colrule \\
 $4S_{1/2}$   & 146.576 &        &   46.148  &       &   33.0457  &        & 6.707  &        &$ -$2.920  & \\
 $4P_{1/2}$   & 16.341  &        &   5.500   &       &   4.820    &        & 0.900  &        &$ -$0.190  & \\
 $4P_{3/2}$   & 3.186   & 22.662 &   1.062   & 7.562 &   1.430    & 13.386 & 0.275  & 1.280  &$ -$0.040  & $-$ 0.305 \\
 $3D_{3/2}$   & 0.464   & 1.168  &   0.430   & 1.080 & $-$0.059    & 2.947  & 0.185  & 0.089  &$ -$0.017  & $-$ 0.090 \\
 $3D_{5/2}$   & 0.199   & 1.666  &   0.183   & 1.527 & $-$0.766    & 4.223  &$-$0.178 & 0.127  &    0.010  & $-$ 0.128 \\
 $5S_{1/2}$   & 38.770  &        &   7.898   &       &   8.200    &        & 0.725  &        &$ -$0.612  &  \\
 $5P_{1/2}$   & 5.683   &        &   1.535   &       &   1.571    &        & 0.100  &        &$ -$0.062  & \\
 $5P_{3/2}$   & 1.089   & 7.752  &   0.296   & 2.114 &   0.463    & 3.967  & 0.068  & 0.206  &$ -$0.013  &  $-$0.096 \\
 $6S_{1/2}$   & 15.043  &        &   2.882   &       &   3.150    &        & 0.246  &        &$ -$0.217  &  \\
 $6P_{1/2}$   & 2.615   &        &   0.633   &       &   0.630    &        & 0.067  &        &$ -$0.027  &  \\
 $6P_{3/2}$   & 0.510   &  3.631 &   0.122   & 0.872 &   0.276    & 1.825  & 0.037  & 0.070  &$ -$0.006  & $-$0.043 \\
 $7S_{1/2}$   & 7.902   &        &   1.148   &       &   1.309    &        & 0.062  &        &$ -$0.104  &  \\
 $8S_{1/2}$   & 4.540   &        &   0.592   &       &   0.746    &        & 0.026  &        &$ -$0.057  &  \\
 $9S_{1/2}$   & 2.687   &        &   0.316   &       &   0.438    &        & 0.011  &        &$ -$0.032  &  \\
 $10S_{1/2}$  & 1.880   &        &   0.013   &       &   0.291    &        & $-$0.014 &        &$ -$0.015  &  \\
\end{tabular}
\end{ruledtabular}
\end{table*}

 We also compare our $A$ results for $^{39}$K obtained using DF, CCSD and CCSD(T) methods with another recent calculations
\cite{safronova} in Table \ref{tab3}. In Ref. \cite{safronova} Safronova and Safronova have also used linearized RCC method 
with singles and doubles approximation (SD method) and including important triples effects for some of the states (SDpT method).
They find large differences between their SD and SDpT results in contrast to our finding of small differences between our CCSD 
and CCSD(T) results. However, both the calculations reveal that the signs of the $A$ values of the $3D_{5/2}$ and $5D_{5/2}$ 
states are negative, which were not resolved correctly in the measurements. Moreover, these calculations indicate that 
correlation effects in the considered atom are substantial for which an all order perturbative method like ours is suitable 
to determine wave functions accurately. To our knowledge, $A$ values are not known experimentally for some of the states 
in $^{39}$K and close agreement between the results from both the calculations in these states will be very useful to conduct 
new measurements in the right direction. Also, we have given $A$ values for few excited states where neither measurements nor 
theoretical calculations are available.
 
\begin{table}
\caption{\label{tab6}
Reported values of nuclear quadrupole moments $Q$ in $b$ for $^{39}K$, $^{40}K$ and $^{41}K$ from various studies.
}
\begin{ruledtabular}
\begin{tabular}{ccc}
\textrm{Isotope}& This work &  Others\\
\colrule \\
$^{39}K$    & 0.0614(6)    & 0.0585 \footnotemark[1] \cite{kello} \\
            &              & 0.0601(15) \cite{olsen} \\
            &              & 0.049(4) \cite{stern}   \\
$^{40}K$    & $-$0.0764(8) &$-$0.073 \footnotemark[1] \cite{kello} \\
            &              & $-$0.0749(19) \cite{olsen}\\
$^{41}K$    & 0.0747(7)    & 0.0711 \footnotemark[1] \cite{kello} \\
            &              & 0.0733(18) \cite{olsen}\\
\end{tabular}
\end{ruledtabular}
\footnote{Accuracy is expected to be better than 1 percent}
\end{table}

Following $A$ results, we present now our calculated $B/Q$ results using the DF, CCSD and CCSD(T) methods
in Table \ref{tab4} for the states where precise experimental $B$ results for $^{39}$K and $^{41}$K are 
available. We also estimate the uncertainties associated with the results obtained using CCSD(T) method 
and present them in the parentheses of the same table. The uncertainties are estimated using the procedure 
as we followed for $A$. To justify that, trends of correlation effects for both the properties behave in a
similar manner, we present contributions from various RCC terms to both $A$ and $B/Q$ results in Table
\ref{tab5} for few important states. It is found that, states where angular momentum is larger than 
half, the contributions from the correlation effects are always of similar scale in magnitude and the
contributions coming from $\overline{O}S_{2v}+cc$ are larger than $\overline{O}S_{1v}+cc$.
This implies that the core-polarization effects are large to estimate $B/Q$ results in the considered
atom which are accounted for, up to all orders through $\overline{O}S_{2v}+cc$ RCC terms in our 
calculations. There are no other calculated results for $B/Q$ available to our knowledge in any of the 
considered isotopes of K to compare with our results.
 
\begin{table*}                                                         
\caption{\label{tab7}
Comparison of estimated and experimental $B$ results in $^{39-41}K$ (in MHz).
}
\begin{ruledtabular}
\begin{tabular}{cccccccc}
\textrm{State} & \textrm{$(B/Q)^{(Th)}$} & \multicolumn{3}{c}{\textrm{This Work}} & \multicolumn{3}{c}{\textrm{Experiments}} \\
  \cline{3-5} \cline{6-8} \\  &     &  $^{39}$K  & $^{40}$K & $^{41}$K  &  $^{39}$K  & $^{40}$K & $^{41}$K \\
\colrule \\
$4P_{3/2}$    & 44.6(5)  &  2.738(41)     &  $-$3.41(5)   & 3.332(49) &  2.786(71)\cite{flake}     &$-$3.445(90) \cite{flake} & 3.351(71)\cite{flake} \\
              &          &                &               &           &  2.9(2)  \cite{schmieder}  &$-$3.23(50)\cite{bendali} & 3.34(24) \cite{ney}   \\    
              &          &                &               &           &  2.72(12) \cite{ney}       &$-$3.5(5)\cite{ney2}     & 3.320(23)\cite{sieradzan}\\ 
              &          &                &               &           &  2.83(13)  \cite{Arimondo} &                      &            \\
$3D_{3/2}$    & 5.2(5)   &  0.319(31)     & $ -$0.40(4)   & 0.388(38) &  0.37(8) \cite{Yei}        &  0.4(1) \cite{Yei}   &  0.51(8)\cite{Yei} \\
$3D_{5/2}$    & 7.4(4)   & \bf{ 0.454(25)}& $ -$0.565(31) & \bf{0.553(30)} & $<$0.3 \cite{Yei}  & 0.8(8)\cite{Yei}    & $<$ 0.2 \cite{Yei} \\     
$4D_{3/2}$    & 2.35(4)  & 0.144(3)       & $ -$0.180(4)  &  0.176(3) &   &     & \\
$4D_{5/2}$    & 3.35(4)  & 0.206(3)       & $ -$0.256(4)  &  0.250(4) &   &     & \\
$5P_{3/2}$    & 13.9(4)  & 0.853(26)      & $ -$1.06(3)   &  1.038(31)& 0.870(22) \cite{schmieder,ney}  &$-$1.16(22)\cite{ney2} & 1.06(4) \cite{ney} \\
              &          &                &               &           & 0.92(10)\cite{schmieder}  &     &\\
$5D_{3/2}$    & 1.16(8)  & 0.071(5)       &  $-$0.088(6)  &  0.087(6) &   &     & \\
$5D_{5/2}$    & 1.65(10) & 0.101(6)       &  $-$0.126(8)  &  0.123(8) &   &     & \\
$6P_{3/2}$    & 6.4(5)   & 0.393(31)      &  $-$0.489(39) &  0.478(38)& 0.370(15)\cite{svanberg}  &     & \\
$6D_{3/2}$    & 0.72(5)  & 0.044(3)       &  $-$0.055(4)  &  0.054(4) & 0.05(2) \cite{glodz}  &     & \\
$6D_{5/2}$    & 1.02(6)  & 0.063(4)       &  $-$0.078(5)  &  0.076(5) &   &     & \\
$7P_{3/2}$    & 3.4(3)   & 0.209(19)      &  $-$0.260(23) &  0.254(23)&   &     & \\
$7D_{3/2}$    & 0.44(3)  & 0.027(2)       &  $-$0.034(2)  &  0.033(2) &   &     & \\
$7D_{5/2}$    & 0.62(4)  & 0.038(2)       &  $-$0.047(3)  &  0.046(3) &   &     &  \\
$8P_{3/2}$    & 2.0(2)   & 0.123(12)      &  $-$0.153(15) &  0.149(15)&   &     & \\
$8D_{3/2}$    & 0.29(2)  & 0.018(1)       &  $-$0.022(2)  &  0.022(2) &   &     &  \\
$8D_{5/2}$    & 0.36(2)  & 0.022(1)       &  $-$0.028(2)  &  0.027(2) &   &     &  \\
$9P_{3/2}$    & 1.5(2)   & 0.092(12)      &  $-$0.115(15) &  0.112(15)&   &     &  \\
$9D_{3/2}$    & 0.92(6)  & 0.056(4)       &  $-$0.070(5)  &  0.069(5) &   &     &  \\
$9D_{5/2}$    & 1.35(8)  & 0.083(5)       &  $-$0.103(6)  &  0.101(6) &   &     &  \\
\end{tabular}                           
\end{ruledtabular}                      
\end{table*}

Both the calculated results for $A$ and $B/Q$ seem to be very accurate, moreover $A$ values are in
good agreement with the available experimental results. There are also several experimental results
available for $B$ in $^{39}$K as well as $^{41}$K which are given in Table \ref{tab4} \cite{flake, 
Arimondo, Yei, banerjee, risberg, schmieder}. The most precise values are quoted in bold fonts for the 
respective states in the same table and we combine these results with our calculated $B/Q$ values to 
estimate $Q$s in these two isotopes. We obtain three different values of $Q$ in $^{39}$K and two values 
in $^{41}$K. All these estimated values agree with each other in their respective uncertainties, but 
the most precise results which are obtained from the $4P_{3/2}$ state are 0.0625(17) $b$ and 0.0744(10) 
$b$ for $^{39}$K and $^{41}$K, respectively. Here we have used the following expression to evaluate the 
net uncertainties of $Q$ values
\begin{eqnarray}
\delta C = C \ \sqrt{\left (\frac{\delta A}{A} \right )^2 + \left (\frac{\delta B}{B} \right )^2},
\end{eqnarray}
where we assume $C$ is the extracted value from $A/B$ and $\delta A$, $\delta B$ and $\delta C$ are their 
respective uncertainties.

Among both the new $Q$ values in $^{39}$K and $^{41}$K, the relative uncertainty in $Q$ of $^{41}$K is
small. Moreover there are also experimental results for the ratios of $Q$ values
between $^{39}$K, $^{40}$K and $^{41}$K are available as \cite{hughes,stern,jones}
\begin{equation}                                                      
\frac{Q(^{40}K)}{Q(^{39}K)}= - 1.244\pm 0.002 
\end{equation}
and 
\begin{eqnarray}
\frac{Q( ^{41}K)}{Q(^{39}K)} = 1.2173 \pm 0.0001 .
\end{eqnarray}   
Using the measured $\frac{Q( ^{41}K)}{Q(^{39}K)}$ and $Q$ value of $^{41}$K, we get a new $Q$ 
value for $^{39}$K as 0.0612(8) $b$. Considering both the values of $Q$ in $^{39}$K,
we restrict the lower and upper limits of $Q( \ ^{39}\text{K})$ to  0.0608 $b$ and
0.0620 $b$, respectively. Therefore, we recommend $Q$ value of $^{39}$K as 0.0614(6) $b$.
Now with this most precise $Q$ value and the above ratios of $Q$ values between different isotopes,
we get precise $Q$ values for  $^{40}$K and $^{41}$K as $-$0.0764(8) $b$ and 0.0747(7) $b$, respectively.
There are also other reported $Q$ values which we have compared them with ours in Table \ref{tab6}.
As seen from the table, there are three other works report these results \cite{kello,
olsen, stern}. Apart from Ref. \cite{stern}, the calculations carried out in these 
works are rigorous and results reported in \cite{kello} are the latest. $Q$ values reported
in \cite{olsen} matches with our estimated values with some overlaps within the predicted uncertainties,
however results reported in \cite{kello} disagree with us. In both these theoretical
works, they have determined electric field gradients at the nucleus to extract the nuclear
quadrupole moments and the results are model independent. But both the calculations are 
less rigorous than the present calculations. In Ref. \cite{olsen}, Sundholm and Olsen have
used a non-relativistic large scale finite-element multi-configuration Hartree-Fock configuration 
interaction (MCHF) method. The core contributions were estimated from the core-valence correlation
calculations and the relativistic corrections are accounted separately from the DF calculation.
Contrast to this work, we have considered the core correlation and core-valence correlations 
to all orders and relativistic effects are included to all orders through the RCC method. In fact,
their truncative CI method is known to have size-consistent problem \cite{szabo} against our RCC 
method. On the other-hand, level of approximations employed to carry out calculations in Ref. 
\cite{kello} by Kell\"o and Sadlej are comparable to the present work. In their work, calculations
are performed with a scalar relativistic Hamiltonian using Douglas-Kroll approach and CCSD(T)
method is used to account the correlation effects. Since this approach is better than the above
MHCF method, the results reported in \cite{kello} were considered to be more accurate and estimated
to be within 1\% accuracy. However, we find the reported values in \cite{kello} are smaller in magnitudes
as compared to our estimations.

 It will be interesting to see further theoretical studies of $B/Q$ results to draw comparison with the
present work. Further, we combine our calculated $B/Q$ results with the newly obtained $Q$ values to
determine theoretical results for $B$ in many states of the considered isotopes of K. These results are
given in Table \ref{tab7}. We have also neglected here the anomalous effects in calculated $B/Q$ values
for different isotopes due to their negligible roles. If some of these $B$ results can be measured more 
precisely than the reported results, then combining those results with our calculated $B/Q$ values will 
definitely give rise better $Q$ values in these isotopes. In fact, more precise theoretical calculations 
of $B/Q$ results in these isotopes can also give rise to more accurate $Q$ values in K.

As can be noticed in Table \ref{tab7}, our estimated $B$ results for different states in K isotopes
agree very well in most of the states except with the $3D$ states. We suggest to carry out further 
measurements of $B$ in these states to ascertain our results. Theoretical results for $B$ are also 
given in many excited states for the first time which can be verified by the future measurements. 

\section{Conclusion}

We have employed relativistic coupled-cluster method to calculate matrix elements of the
hyperfine interaction Hamiltonians in potassium atom. By performing calculations of the magnetic dipole 
hyperfine structure constants in this atom, we have tested the accuracies of the wave functions in the 
nuclear region. These wave functions were further used for the electric quadrupole hyperfine interaction 
studies. By combining our calculations with the corresponding measurements, we obtained the nuclear 
quadrupole moments as $0.0614(6) \ b$, $-0.0764(8) \ b$ and $0.0747(7) \ b$ for $^{39}$K, $^{40}$K and 
$^{41}$K, respectively. These results agree with one of the previous work but do not agree with others 
including the latest reported results. After obtaining nuclear quadrupole moments, we substituted them 
to obtain electric quadrupole hyperfine structure constants in many states and found very good agreement
with the experimental results except for the $3D$ states. Also, we have given some of the results that 
were not reported earlier. We suggest further studies of the considered properties to ascertain our 
results.

\section{Acknowledgment}
Computations were carried out using 3TFLOP HPC cluster of Physical Research Laboratory, Ahmedabad.

\end{document}